\documentclass[%
 superscriptaddress,
 nofootinbib,
 amsmath,amssymb,
 graphicx,
 aps, physrev,
 numbers,super
]{revtex4-2}
\usepackage{graphicx}
\usepackage{tabularx}
\usepackage{bm} 
\usepackage[version=4]{mhchem}
\usepackage[section]{placeins} 
\usepackage{siunitx}
\usepackage{pdftexcmds}
\usepackage{microtype}
\usepackage{hyperref}
\usepackage[capitalise]{cleveref} 

\usepackage[format=plain,justification=justified]{caption} 
\newcommand{\sub}[1]{$_{\text{#1}}$}
\makeatletter
\newcommand\kv[2]{%
  \ifnum\pdf@strcmp{\unexpanded{#1}}{V}=0 %
     \expandafter\@firstoftwo
  \else
    \expandafter\@secondoftwo
  \fi
    {\textit{#1}\!\sub{#2}}
    {#1\sub{#2}}%
}
\makeatother
\makeatletter
\newcommand\kvc[3]{%
  \ifnum\pdf@strcmp{\unexpanded{#1}}{V}=0 %
     \expandafter\@firstoftwo
  \else
    \expandafter\@secondoftwo
  \fi
    {\textit{#1}\!\sub{#2}$^{#3}$}
    {#1\sub{#2}$^{#3}$}
}
\makeatother

\DeclareSIUnit\angstrom{\protect \text {Å}}

\hypersetup{
    colorlinks=true,                
    breaklinks=true,                
    urlcolor= black,                
    linkcolor= blue,                
    bookmarksopen=false,
    filecolor=black,
    citecolor=magenta,
}

\newcommand{\angstrom}{\mbox{\normalfont\AA}} 
\DeclareSIUnit\angstrom{\text {Å}}

\begin{document}
\preprint{APS/123-QED}

\title{Dynamic Vacancy Levels in \ce{CsPbCl3} Obey Equilibrium Defect Thermodynamics}

\author{Irea Mosquera-Lois}
\affiliation{Thomas Young Centre \& Department of Materials, Imperial College London, London SW7 2AZ, UK}
\author{Aron Walsh}
\email{a.walsh@imperial.ac.uk}
\affiliation{Thomas Young Centre \& Department of Materials, Imperial College London, London SW7 2AZ, UK}

\date{\today}
             
\begin{abstract} 
Halide vacancies are the dominant point defects in perovskites with \kv{V}{Cl} identified as a detrimental trap for the optoelectronic performance of \ce{CsPbCl3}, with applications ranging from photodetectors to solar cells. Understanding these defects under operating conditions is key since their electronic levels exhibit large thermal fluctuations that challenge the validity of static 0 K models. However, quantitative modelling of defect processes requires hybrid density functional theory with spin-orbit coupling, which is too expensive for direct molecular dynamic simulations. To address this, we train a multi-task machine learning force field to study \kv{V}{Cl} in orthorhombic \ce{CsPbCl3} at 300 K.While we observe strong oscillations in the optical transition level arising from the soft potential energy surface, neither the non-radiative capture barriers nor the thermodynamic charge transition levels are affected. Our results reveal that \kv{V}{Cl} is not responsible for the non-radiative losses previously assumed. Instead, its impact on performance arises from other mechanisms, such as limiting the open-circuit voltage and promoting ionic migration. Our findings demonstrate that, despite strong dynamical effects in halide perovskites, the conventional static formalism of defect theory remains valid for predicting thermodynamic behavior, providing a sound basis for the design of high-performance energy materials.
\end{abstract}


\maketitle

\section{Introduction}

Point defects dictate the physical properties of most functional materials. Their dilute concentrations in crystals challenge experimental characterisation, requiring a combination of experiment and theory to understand and optimise defect behavior. Typically, theoretical studies assume a static framework by considering the \SI{0}{K} ground-state structures for all possible defect species and calculating their concentrations and optoelectronic properties\cite{mosquera-lois_imperfections_2023}. However, recent studies have questioned the validity of this approximation for soft crystals like metal halide perovskites, which exhibit large-amplitude and anharmonic structural dynamics at room temperature\cite{egger_what_2018,zacharias2023anharmonic,dubajic2025dynamic,PhysRevMaterials.7.L092401}. It remains an open question whether this is linked to the empirical defect tolerance observed for perovskite materials and devices\cite{mosquera-lois_multifaceted_2025}.

There have been multiple reports on the strong and slow fluctuations of the electronic levels of halide vacancies in \ce{CsPbBr3}\cite{cohen_breakdown_2019,ran_halide_2023,deng_strong_2024}, \ce{CsPbI3}\cite{chu_soft_2020,liu_significant_2024,shi_band_2025} and \ce{CH3NH3PbI3}\cite{chu_low-frequency_2020,li_control_2018,wang_electron-volt_2022} at device operating conditions of solar cells, resulting in the defect oscillating between deep to a shallow character in the picosecond timescale, and thus drastically changing its behavior. However, the conclusions of these reports are limited by their level of theory (simpler approximations within density functional theory), which can lead to an incorrect description of the potential energy surface (PES) for the halide vacancies\cite{kang_effects_2020,ming_defect_2022,zhang_iodine_2023,kang_atomic-scale_2021,meggiolaro_iodine_2018,meggiolaro_first-principles_2018,du_density_2015,zhang_defect_2022} and overestimates the defect level fluctuations\cite{kang_effects_2020}. Another limitation stems from their focus on the variation of electronic \emph{eigenvalues} and optical levels rather than the \emph{thermodynamic} transition levels.
While optical or vertical levels are important for radiative processes (e.g., defect absorption or photo-luminescence), the impact of defects on device performance is often controlled by non-radiative processes like non-radiative carrier capture, which is determined by the thermodynamic transition levels\cite{zhang_iodine_2023,mosquera-lois_multifaceted_2025}. 

In this study, we overcome these limitations by using an accurate level of theory to describe defect processes in lead halide perovskites that combines the hybrid exchange-correlation functional (HSE\cite{Heyd2003}) with spin-orbit coupling (SOC). Using this description, we quantify the fluctuations of the optical transition levels at \SI{300}{K} and relate them to the corresponding thermodynamic charge transition levels, thereby revealing how thermal motion affects defect electronic behavior in \ce{CsPbCl3}.
Due to the high computational cost of HSE+SOC, we leverage multi-task machine learning force fields to run molecular dynamics over long time and length scales that would be prohibitively expensive from first-principles. 
We focus on the chloride vacancy (\kvc{V}{Cl}{0,+1}) in orthorhombic\cite{hirotsu_experimental_1971,cohen_phase_1971,haeger_thermal_2020,he_demonstration_2021} \ce{CsPbCl3}, since it has been suggested as the main defect promoting non-radiative carrier recombination and damaging device performance\cite{dongre_s_dual_2023,chen_roles_2024,sommer_defect_2022,ma_insights_2019,zhu_recent_2021,seth_tackling_2019,fiuza-maneiro_unlocking_2025,zheng_reducing_2019,fei_unlocking_2025,peters_defect_2022}, thereby motivating a range of passivating strategies\cite{ye_defect_2021,cheng_passivating_2022,dongre_s_dual_2023,zhang_engineering_2021,ahmed_giant_2018,yong_doping-enhanced_2018,li_simultaneous_2021,jiang_ultrafast_2023,fiuza-maneiro_unlocking_2025}. 

\section{Results}

\subsection{Machine learning force field}

\begin{figure}[ht!]
  \centering
  \includegraphics[width=0.9\columnwidth]{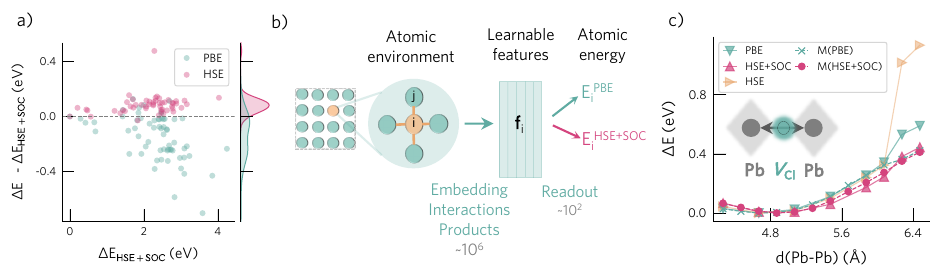}
  \caption{
  a) Errors in the relative energies of \kvc{V}{Cl}{0} for a range of sampled configurations from molecular dynamics between $0-300$~K calculated with PBE (green) and HSE (pink) in comparison with HSE+SOC. Both PBE and HSE lead to significant differences and will thus affect the resulting dynamics.
  b) Schematic of the architecture of the multihead model, illustrating the common operations and parameters to learn the atomic features and the separate readout layers used for each level of theory. 
  c) Potential energy surface of \kvc{V}{Cl}{0} (at the site with $C_s$ symmetry) in orthorhombic \ce{CsPbCl3}, mapping the energy versus the distance between the Pb atoms neighbouring the vacancy. The reference DFT data is shown with triangles, with different colours indicating the different levels of theory. 
  Model predictions (dashed lines with crosses (M(PBE)) and circles (M(HSE+SOC)) accurately reproduce the reference data. 
  }
  \label{fig:pes}
\end{figure}

To accurately describe the PES of halide vacancies in lead-based perovskites, a high level of theory is required\cite{kang_effects_2020,ming_defect_2022,zhang_iodine_2023,kang_atomic-scale_2021,meggiolaro_iodine_2018,meggiolaro_first-principles_2018,du_density_2015}. 
Relativistic effects are necessary to describe the Pb 6p-based conduction band, which results in an energy shift of almost \SI{1}{eV} when spin-orbit coupling is included.\cite{brivio2014relativistic,umari2014relativistic}
The use of a hybrid exchange-correlation functional is further necessary to remove large errors in the band gap and the alignment of defect levels with the band edges.\cite{du_density_2015,meggiolaro_first-principles_2018} 
As illustrated in \ref{fig:pes}a, a low-level of theory like PBE with scalar-relativistic effects leads to significant errors in the relative energies of \kvc{V}{Cl}{0} when compared to a high-level reference (HSE+SOC), especially for high-energy configurations. While a hybrid functional (HSE) reduces these errors, it is still limited for configurations with a long separation between the vacancy nearest neighbours (\ref{fig:pes}c), thus resulting in incorrect dynamics and the need to include SOC. However, the computational cost of advanced methods like HSE+SOC makes them impractical for molecular dynamics simulations to probe defect behavior at finite temperatures. 

To overcome this obstacle, we use a multihead MACE\cite{mace} machine learning force field that integrates training data calculated with two levels of theory. This approach allows us to leverage a large dataset of low-cost, low-fidelity calculations (PBE), supplemented by a much smaller set of expensive, high-fidelity data (HSE+SOC). As shown in \ref{fig:pes}b, the model shares most of its parameters for transforming atomic environments into features but uses separate readout layers to map these features to atomic energies. This architecture enables the model to learn general structure–energy relationships from the low-fidelity data, while the high-fidelity data corrects for the subtle differences between the two potential energy surfaces. 
The trained model can reproduce both levels of theory and accurately describes the PES of \kvc{V}{Cl}{0} as the distance between the Pb atoms neighbouring the vacancy varies (\ref{fig:pes}c). 
Finally, to describe the transition between charge states, we require an accurate description of the potential energy surface for both \kvc{V}{Cl}{0} and \kvc{V}{Cl}{+1}. We therefore train and validate separate models for each charge state, with both models achieving good accuracies on their respective test sets (see Appendix A).

\subsection{Optical defect level}

\begin{figure}[ht!]
  \centering
  \includegraphics[width=0.8\textwidth]{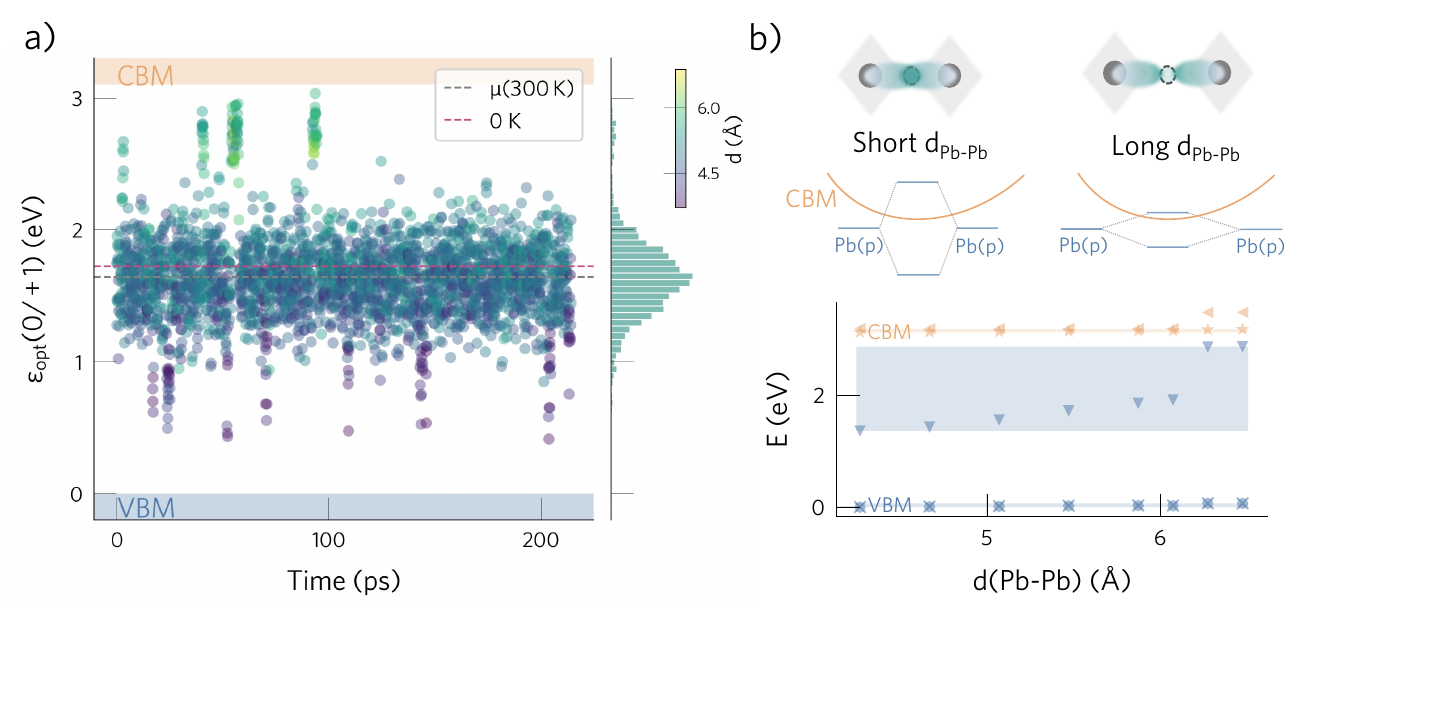}
  \caption{
  a) Variations of the optical defect level $\epsilon_\mathrm{opt}(0\rightarrow+1)$ at \SI{300}{K} and comparison with the \SI{0}{K} value (pink) for a 640 atom supercell containing a single vacancy. The colormap quantifies the separation between the Pb neighbouring the vacancy, with shorter distances resulting in stronger Pb(p) overlap and thus a level deeper in the gap. 
  b) Electronic eigenvalues (calculated with HSE+SOC) of \kvc{V}{Cl}{0} configurations in a 160-atom supercell with increasing $\mathrm{Pb-Pb}$ distance, illustrating that the defect level approaches the conduction band minimum as the separation between Pb atoms increases, as previously reported in the literature\cite{cohen_breakdown_2019}. Occupied and unoccupied levels are shown in blue and orange, respectively, while the filled area highlights the variation of each eigenvalue. 
  }
  \label{fig:optical}
\end{figure}

After validating the models, we apply them to investigate how thermal effects affect the optical defect level, which corresponds with an instantaneous transition between charge states where the defect geometry remains fixed.  
To evaluate the optical level $\epsilon_\mathrm{opt}^{0\rightarrow+1}$ at a typical operating temperature of \SI{300}{K}, we perform NPT molecular dynamics for the neutral charge state using the model trained on \kvc{V}{Cl}{0}. For each frame $R$ in the neutral trajectory, we calculate the energy difference between the neutral and positively-charged states using separate machine learning models for each charge state:
\begin{equation}
\epsilon_\mathrm{opt}^{0\rightarrow+1}(R_{0}) = E^{+1}(R_{0}) - E^{0}(R_{0}),
\end{equation}
where the energies $E^{q=0}$ and $E^{q=+1}$ are evaluated with the models for the neutral and positively charged vacancy, respectively, and $R_0$ denotes the structure in the neutral state. The time evolution of $\epsilon_\mathrm{opt}^{0\rightarrow+1}$ exhibits strong fluctuations of more than \SI{1}{eV} (\ref{fig:optical}a), which agrees with previous reports on halide vacancies in \ce{CsPbBr3}\cite{cohen_breakdown_2019} and \ce{CsPbI3}\cite{chu_soft_2020,liu_significant_2024,shi_band_2025} and observations of broad sub-gap emission around \SI{1.65}{eV} in photoluminescence spectra of \ce{CH3NH3PbCl3}\cite{wang_fast_2023,cheng_passivating_2022}. 

The fluctuations of $\epsilon_\mathrm{opt}^{0\rightarrow+1}$ are partly caused by changes in the separation between the two Pb atoms neighbouring the vacancy. When a neutral Cl vacancy forms, the dangling $p$ orbitals of the Pb atoms neighbouring the vacancy interact and form bonding and antibonding combinations, with the bonding orbital appearing in the band gap and hosting one electron. At finite temperature, the Pb atoms oscillate around their equilibrium position, with shorter $\mathrm{Pb-Pb}$ distances resulting in stronger orbital overlap and thus lowering the energy of the bonding orbital. This relationship is illustrated in \ref{fig:optical}b by evaluating with DFT (HSE+SOC) the electronic eigenvalues for \kvc{V}{Cl}{0} configurations with increasing $\mathrm{Pb-Pb}$ separation. 

Beyond changes in $d(\mathrm{Pb-Pb})$, $\epsilon_\mathrm{opt}^{0\rightarrow+1}$ is also broadened by thermal fluctuations of the global structure, including volume variations and octahedral tilting and distortions, reflecting the softness of the perovskite lattice. 
Despite the significant ionic motion at \SI{300}{K} and its impact on $\epsilon_\mathrm{opt}^{0\rightarrow+1}$, we find that its ensemble average is in good agreement with the static prediction based on the \SI{0}{K} ground-state structure of \kvc{V}{Cl}{0} (1.65 vs 1.58~eV). This agreement supports the validity of the static approximation for predicting the average position of optical levels even in soft materials with strong dynamics.  

\subsection{Effect on non-radiative carrier capture} 
\begin{figure}[ht!]
    \centering
    \includegraphics[width=0.94\linewidth]{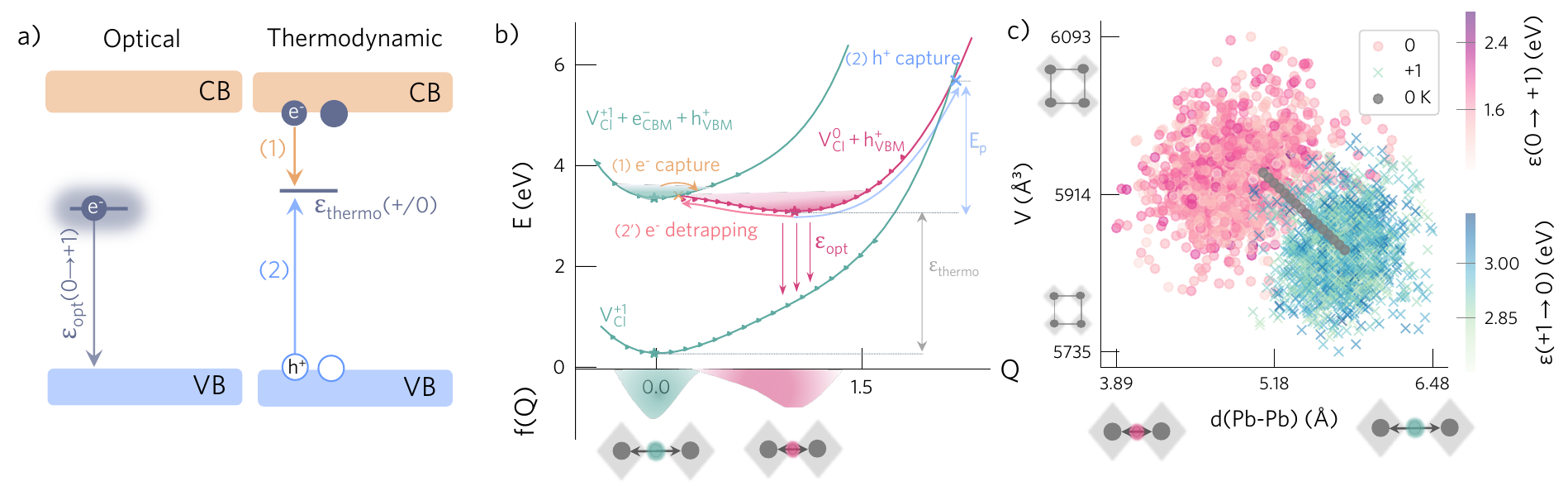}
    \caption{
a) Schematic energy-level diagram illustrating two types of charge transitions. The optical transition level ($\epsilon_\mathrm{opt}$) corresponds to an instantaneous (vertical) change in charge state. In contrast, the thermodynamic transition level ($\epsilon_\mathrm{thermo}$) reflects a thermally activated process in which the defect relaxes between the equilibrium geometries of the two charge states.
b) Configurational coordinate (CC) diagram showing the transition between charge states \kvc{V}{Cl}{+1} and \kvc{V}{Cl}{0} along a generalized reaction coordinate 
$Q$ that tracks structural relaxation. The green and pink parabolas represent the PES of \kvc{V}{Cl}{+1} (before and after photo-excitation) and \kvc{V}{Cl}{0}, respectively. Stars mark the equilibrium configurations of each charge state. Vertical pink arrows indicate possible optical transitions $\epsilon_\mathrm{opt}(0\rightarrow+1)$, while orange and blue arrows depict the adiabatic energy barriers for non-radiative carrier capture. The shaded distributions $f(Q)$ visualize the population of lattice configurations at finite temperature, with their widths reflecting the softness of each PES and the amplitude of vibrational motion around equilibrium.
c) 2D map illustrating the structural variation of the defect in both charge states at \SI{300}{K} (from NPT trajectories of the neutral and positively charged states, represented with pink dots and blue crosses, respectively) and comparison with the 1D interpolated path connecting their equilibrium structures (grey dots). The colorbars illustrate the variation of the optical levels for each configuration in the trajectories. 
}
    \label{fig:cc}
\end{figure}

It has been postulated that the slow and strong variations of $\epsilon_\mathrm{opt}^{0\rightarrow+1}$ can change the level character from shallow to deep and thus change how the defect traps free carriers and its effect on non-radiative carrier capture.\cite{cohen_breakdown_2019} 
However, non-radiative carrier capture is an adiabatic process 
governed by the minimum-energy path between charge states on their respective PES. 
This process is illustrated in the configurational coordinate (CC) diagram of \ref{fig:cc}b, where the PES for each charge state is plotted along a generalized coordinate $Q$ that reflects the structural changes during the transition --- dominated here by the $\mathrm{Pb-Pb}$ distance adjacent to the vacancy, which decreases upon electron capture. 
After photo-excitation generates excess charge carriers, an electron capture event requires the system, initially at the equilibrium geometry of \kvc{V}{Cl}{+1}, to overcome (or tunnel through) the energy barrier. This barrier originates from the intersection of the PES for the two charge states and is not directly related to the vertical optical transition, which corresponds to instantaneous (vertical) excitations. Thus, the variations of $\epsilon_\mathrm{opt}^{0\rightarrow+1}$
reflect possible optical transitions as the defect samples the neutral state's PES and reflect the softness of that surface --- as illustrated in \ref{fig:cc}c which shows the range of possible optical transitions during MD trajectories of the neutral and charged vacancies. This broad $\epsilon_\mathrm{opt}$ is also a consequence of strong electron-phonon coupling\cite{turiansky2025machinelearningphononspectra}, as quantified by the large Huang-Rhys factor of 24.4. However, these variations do not influence the recombination kinetics, which are determined by the capture barriers. 

From the configurational coordinate diagram, we find that electron capture by \kvc{V}{Cl}{0} is a fast process due to the small energy barrier ($E_n = 0.05$~eV). To complete the non-radiative recombination cycle, however, \kvc{V}{Cl}{0} would need to also capture a hole. This process is hindered by a high barrier for non-radiative hole capture ($E_p = 2.53~$eV)
, suggesting that \kv{V}{Cl} primarily acts as a reversible electron trap. This is consistent with experimental observations of fast electronic trapping in \ce{CH3NH3PbCl3} and assigned to \kv{V}{Cl}\cite{wang_fast_2023}.
 
This behavior aligns with trends observed for halide vacancies in other perovskites. For example, \kv{V}{I} in \ce{CsBI3} (B = Pb, Sn, Ge) also exhibits negligible contribution to non-radiative recombination due to high capture barriers\cite{zhang_iodine_2023}. 
While halide vacancies seem inactive as non-radiative recombination centres, they can still limit device performance through other mechanisms. 
In the presence of a high density of ($+/0$) trap states below the conduction band, the electron quasi-Fermi level will become pinned under illumination. As the concentration of halide vacancies is expected to exceed the photo-generated free carrier density ($\approx10^{16}~\mathrm{cm^{-3}}$ under one sun conditions), these trap states will act as an electron reservoir. This will limit the build-up of conduction band electrons, suppressing the quasi-Fermi level splitting and capping the achievable open-circuit voltage in a solar cell, which supports observations of voltage improvements with Cl-rich conditions in \ce{CH3NH3PbCl3}\cite{zia_mapbcl3_2024,zhao_reduced_2023} and \ce{CsPbCl3}\cite{cheng_passivating_2022}. 
Beyond pinning the Fermi level, the presence of \kv{V}{Cl} also promotes ion migration, and enhances local lattice distortions (Figure A5). These factors may underlie the observed performance improvements when the concentration of \kv{V}{Cl} is reduced via (post)synthesis treatments.

\subsection{Thermodynamic defect level} 

Predicting the position of the thermodynamic charge transition level ($\epsilon_\mathrm{thermo}$) is critical to evaluating the relative stability of defect charge states and how they affect device performance. Within the static configurational coordinate framework, $\epsilon_\mathrm{thermo}$ corresponds to the energy difference between the minima of the potential energy surfaces for different charge states, and it determines the equilibrium occupation of defect levels. 
Accurate predictions of $\epsilon_\mathrm{thermo}$ are thus not only essential for linking spectroscopic measurements to specific defects, but also for estimating free carrier concentrations and linking synthesis conditions to device performance. 

The transition level is often predicted under a static 0 K approximation, where the level is given by the difference in internal energies of the defect at the relaxed geometry of each charge state, e.g. $\epsilon_\mathrm{thermo} = E^{0}(R^0) - E^{+1}(R^{+1}) - \epsilon_\mathrm{VBM}$. However, devices operate at finite temperatures, and thus in reality, the level should be calculated from the difference in \emph{Gibbs free energies}, $\epsilon_\mathrm{thermo}(T) = G^{0}(R^0) - G^{+1}(R^{+1}) - \epsilon_\mathrm{VBM}$\cite{qiao_temperature_2022}. To assess the limitations of the 0 K approximation, we compare both frameworks. In our finite-temperature approach, we evaluate the free energies by averaging the total internal energies sampled from NPT molecular dynamics simulations\cite{wu_defects_2022} at \SI{300}{K} and decomposing the entropy into spin and vibrational terms:
\begin{equation}
    \epsilon_\mathrm{thermo}(+1/0) = (\langle E^0 \rangle - TS^0) - ( \langle E^{+1} \rangle -TS^{+1}) - \epsilon_\mathrm{VBM}. 
\end{equation}
where the spin component is determined from the total spin angular momentum $S$, $S_\mathrm{spin}= k_B \ln{(2S + 1)}$, and the vibrational term is evaluated using the harmonic approximation. Note that here we neglect the temperature dependence of the bulk band edges\cite{qiao_temperature_2022} (e.g. $\epsilon_\mathrm{VBM}(T)$) since we are focusing on the intrinsic defect effects (e.g., how defect dynamics affect $\epsilon_\mathrm{thermo}(+1/0)$).

To understand the contribution of different factors, we calculate a series of $\epsilon_\mathrm{thermo}(+1/0)$. In order of increasing sophistication: i) at \SI{0}{K} using the constant volume approach (i.e., the lattice fixed to its bulk equilibrium values), ii) at \SI{0}{K} and constant pressure (i.e., allowing lattice relaxation for each charge state), iii) at \SI{300}{K} but neglecting entropies, and iv) at \SI{300}{K} and including entropies. The first three approaches are in good agreement, with $\epsilon_\mathrm{thermo}$ values of 2.71, 2.76 and 2.79 eV above the VBM, which demonstrates that lattice relaxation and structural motion at \SI{300}{K} only change $\epsilon_\mathrm{thermo}$ by \SI{50}{meV} and \SI{30}{meV}, respectively. The difference in spin and vibrational entropy are $-T\Delta s_\mathrm{spin} (+1\rightarrow0)=$\SI{-18}{meV} and $-T\Delta s_\mathrm{vib} (+1\rightarrow0)=$\SI{-109}{meV} at \SI{300}{K}, respectively. This results in a cancellation of effects between the volume relaxation and the vibrational entropy, resulting in good agreement between the static \SI{0}{K} and the dynamic \SI{300}{K} values (with transition levels of \SI{2.71}{eV} and \SI{2.66}{eV}, respectively). This analysis validates the accuracy of the static \SI{0}{K} approximation for predicting charge transition levels for vacancies in halide perovskites. 

\section{Discussion and Conclusions}

A longstanding question in the defect physics of halide perovskites has been whether their softness and dynamic disorder undermine the conventional, static framework used to describe defects in semiconductors. Our results show that this is not the case. We have investigated the dynamics of defect levels in halide perovskites, addressing directly whether the static picture of defect energetics remains valid for these materials. By performing molecular dynamics simulations, we have confirmed previous reports of large fluctuations in the optical level of halide vacancies, which persist even at the HSE+SOC level of theory. These fluctuations originate from large-amplitude thermal vibrations, a consequence of the anharmonic potential energy surfaces of halide vacancies. However, our analysis clarifies that these dynamic variations do not alter thermodynamic transition levels or non-radiative recombination kinetics. In fact, such dynamics are a natural manifestation of the defect motion that traditional configurational coordinate diagrams are designed to represent.
These thermally-activated processes involve slow adiabatic transitions and are governed by energy barriers defined by the crossing of potential energy surfaces of different charge states.

Beyond the impact of the level oscillations, we have also validated the accuracy of the static \SI{0}{K} approximation for predicting optical and thermodynamic transition levels for soft and dynamic crystals like halide perovskites. By comparing the properties calculated at \SI{0}{K} and \SI{300}{K}, we have found the static approximation to give accurate predictions, thereby reducing computational costs without sacrificing accuracy. 

By studying the potential energy surfaces of \kv{V}{Cl}, we have found it to be benign for non-radiative recombination, following similar trends to other halide vacancies in \ce{CsBI3} (B=Pb, Sn, Ge) and reflecting their defect tolerance. While the kinetics are too slow for non-radiative recombination, \kv{V}{Cl} can still limit performance by pinning the electron quasi-Fermi level and promoting ionic migration, explaining why Cl-rich conditions result in improved performance.   

Finally, we have illustrated how to calculate optical and thermodynamic defect levels at finite temperatures from molecular dynamics simulations using machine learning force fields, extending the efforts of applying MLFFs to characterise defect behavior\cite{mosquera-lois_point_2025,turiansky2025machinelearningphononspectra,Kavanagh2025}. This approach is useful for materials that adopt a dynamically stable phase at the device operating conditions (e.g., perovskites), where the defect structure and levels may vary with the bulk phase. 
Further, we have demonstrated that multi-task machine learning force fields reduce the cost of generating training data, thereby enabling long molecular dynamics simulations at a high level of theory that would be prohibitive from direct first-principles or a standard MLFF. 
Beyond point defects, this framework enables higher accuracies to study systems with complex configurational landscapes or where simple exchange-correlation functionals fail to provide a reliable description. This includes materials with heavy elements or strong electron correlation, catalytic surfaces where electron localisation is key or subtle phase stabilities of polymorphic materials. 

\section{Methods}

\emph{\textbf{Density Functional Theory calculations.}} All reference total energy and force calculations were performed with Density Functional Theory using the projector augmented wave method\cite{Kresse_1996}, as implemented in the Vienna Ab initio Simulation Package (\texttt{VASP})\cite{Kresse_1993,Kresse_1994}. We used the recommended PBE projector-augmented wave potentials (version 64) for Cs ($5s^2 5p^6 6s^1$), Pb ($5d^{10} 6s^2 6p^2$) and Cl ($3s^2 3p^5$). We used two different exchange-correlation functionals to generate two separate datasets with PBE\cite{perdew_generalized_1996} and HSE+SOC\cite{Heyd2003,steiner_calculation_2016}. The mixing and screening parameters in the HSE functional were set to $\alpha=0.375$ and $0.1$ \AA$^{-1}$ in order to reproduce the bandgap of \ce{CsPbCl3} measured at \SI{300}{K}, as previously done in the literature\cite{lyons_trends_2023}. We used an energy cut-off for the plane-wave basis set of \SI{400}{eV} and a reciprocal space grid of $1\times1\times2$ for the 180-atom supercell. 

\emph{\textbf{Training of machine learning force fields.}}
We used the structure similarity kernel in \texttt{VASP} to generate the PBE training sets of configurations using its on-the-fly molecular dynamics approach\cite{jinnouchi_phase_2019,jinnouchi_--fly_2019,jinnouchi_--fly_2020,liu_phase_2022}. This involved heating runs performed under the NPT ensemble with a pressure of \SI{1}{atm} and from an initial temperature of \SI{100}{K} up to 30\% above our target temperature of 300 K. This procedure was repeated for bulk \ce{CsPbCl3} and the two charge states of \kv{V}{Cl}. We then used the DIRECT method to select the 10\% most diverse structures in each dataset (bulk, \kvc{V}{Cl}{0} and \kvc{V}{Cl}{+1}) and evaluate them at the HSE+SOC level, resulting in the initial high-fidelity datasets.

After generating the initial training sets with \texttt{VASP}, we trained two {\color{black}separate} \texttt{MACE}\cite{mace} force fields for \kvc{V}{Cl}{0} and \kvc{V}{Cl}{+1} on the bulk+defect datasets to obtain models with higher accuracy and speed. 10\% of the configurations in these datasets were used as validation sets to monitor the loss during training.
We used a \texttt{ScaleShiftMACE} model with Ziegler-Biersack-Littmark (ZBL) pair repulsion\cite{Ziegler1985}, 2 message passing layers, 128 equivariant messages, correlation order of 3, angular resolution of 3 and cutoff radius of \SI{5.5}{\angstrom}. 
The batch size was set to 2 and the Huber loss function was used, with weights of 1, 100 and 100 for the mean square errors in the energies, forces and stresses, respectively. 
For the last 20\% of the training epochs, the weights were updated to values of 1000, 10 and 100 for energy, force and stress, respectively --- following the recommended strategy of increasing the weight on the energy errors during the final training epochs. The models were trained until the validation loss converged, which required around 150 epochs. The reference energies were defined as the potential energies of isolated Cs, Pb, and Cl atoms. We used these trained models to run molecular dynamics at 300 K, sample the 10 most diverse configurations, compute their errors via DFT, add the resulting data to our training sets, and retrain the models---repeating this loop until the models achieved our target accuracy.

\emph{\textbf{Point defect calculations.}} The reference DFT defect calculations were setup and analysed using \texttt{doped}\cite{doped}. To account for spurious finite-size supercell effects, the Kumagai-Oba\cite{kumagai_electrostatics-based_2014} (eFNV) charge correction scheme was used to calculate $E_{corr}$, as automated in \texttt{doped}. This correction was calculated for the lowest energy (static) defect configuration and applied a posteriori to the machine learning force field predictions for the total energies of the charged defects. We note that this is an approximation, as the correction (effective screening) should be configuration dependent; however, there is no general solution for charged defects. The low defect charge and large dielectric screening ensure that this term is small (60 meV for the 80-atom supercell). 
To identify the defect ground-state structures at \SI{0}{K}, we used the ShakeNBreak method\cite{mosquera-lois_search_2021,mosquera-lois_identifying_2023,mosquera-lois_shakenbreak_2022}. Carrier capture calculations were performed with the \texttt{CarrierCapture} code\cite{kim_carriercapturejl_2020} using a 80-atom supercell of the orthorhombic structure ($l=16,16,11$\AA) and reciprocal space grid of $1\times1\times2$. 

\emph{\textbf{Validation of machine learning force fields.}}
To generate the test sets, we performed NPT molecular dynamics simulations with the trained models at \SI{300}{K}, running three independent \SI{40}{ps} runs. We then sampled 30 diverse configurations from these trajectories, and performed DFT calculations on them, which were used to calculate the MAE and RMSE on the predicted properties (energy, forces and stresses) of each model (see errors in SI). 
The MAE and RMSE for the forces and stresses were calculated component wise, as defined in Ref.\citenum{morrow_how_2023}. The predicted properties exhibited small absolute errors with no outliers, confirming that the models accurately describe the potential energy surfaces of each defect at \SI{300}{K}. We also validated the models on the transition between charge states by performing a linear interpolation between the equilibrium structures in each charge and comparing the predicted and reference energies. 

\emph{\textbf{Molecular dynamics.}}
To model the behavior of the defects at room temperature, we performed NPT molecular dynamics with \texttt{LAMMPS}\cite{LAMMPS} using both a 160-atom ($l=16, 16, 22$~\AA) and 640-atom supercells ($l=31, 31, 22$~\AA). The Nos\'e-Hoover thermostat and barostat were used (1 atm, 300~K and timestep of \SI{3}{fs}). 
The \SI{300}{K} optical level was evaluated after equilibration and production times of \SI{50}{ps} and \SI{200}{ps}, respectively. 
The vibrational entropies of \kvc{V}{Cl}{0} and \kvc{V}{Cl}{+1} were calculated with \texttt{phonopy} using a 160-atom supercell. 
These trajectories were analysed with Python using tools from the \texttt{ase}\cite{ase-paper}, \texttt{pymatgen}\cite{Ong2013,Jain2013,Ong2015}, \texttt{dscribe}\cite{dscribe,dscribe2}, \texttt{umap}\cite{mcinnes_umap_2018}, \texttt{direct}\cite{qi_robust_2024},
\texttt{matplotlib}\cite{Hunter:2007}, 
and \texttt{seaborn}\cite{Waskom2021} packages, and visualised with \texttt{Ovito}\cite{ovito} and \texttt{CrystalMaker}\cite{crystalmaker}.\\

\section*{Declarations}
{\bf Data availability.}
The datasets and trained models will be available from a Zenodo repository upon publication.

\begin{acknowledgments}
We thank Xinwei Wang for initial calculations on the electronic structure and carrier capture behavior of \kv{V}{Cl} in \ce{CsPbCl3}. 
I.M.-L. thanks Imperial College London (ICL) for funding a President's PhD scholarship. 
We are grateful to the UK Materials
and Molecular Modelling Hub for computational resources, which is partially funded by EPSRC (EP/P020194/1 and EP/T022213/1). This work used the ARCHER2 UK National
Supercomputing Service (https://www.archer2.ac.uk) via our membership of the UK’s HEC Materials Chemistry Consortium, funded by EPSRC (EP/L000202). We acknowledge the ICL High Performance Computing services for computational resources. 
\end{acknowledgments}

{\bf Author contributions.}
I.M.-L.: Investigation, methodology and formal analysis.
A. W.: Conceptualisation \& project administration.
Supervision: A.W. Writing - original draft: I.M.-L. Writing - review \& editing: Both authors. These author contributions are defined according to the CRediT contributor roles taxonomy.

{\bf Competing interests.}
The authors declare no competing interests.

\appendix
\renewcommand\thefigure{A\arabic{figure}}    \setcounter{figure}{0}

\section{\label{app:subsec}Validation of machine learning force fields}

\begin{figure}[ht]
    \centering
    \includegraphics[width=0.9\linewidth]{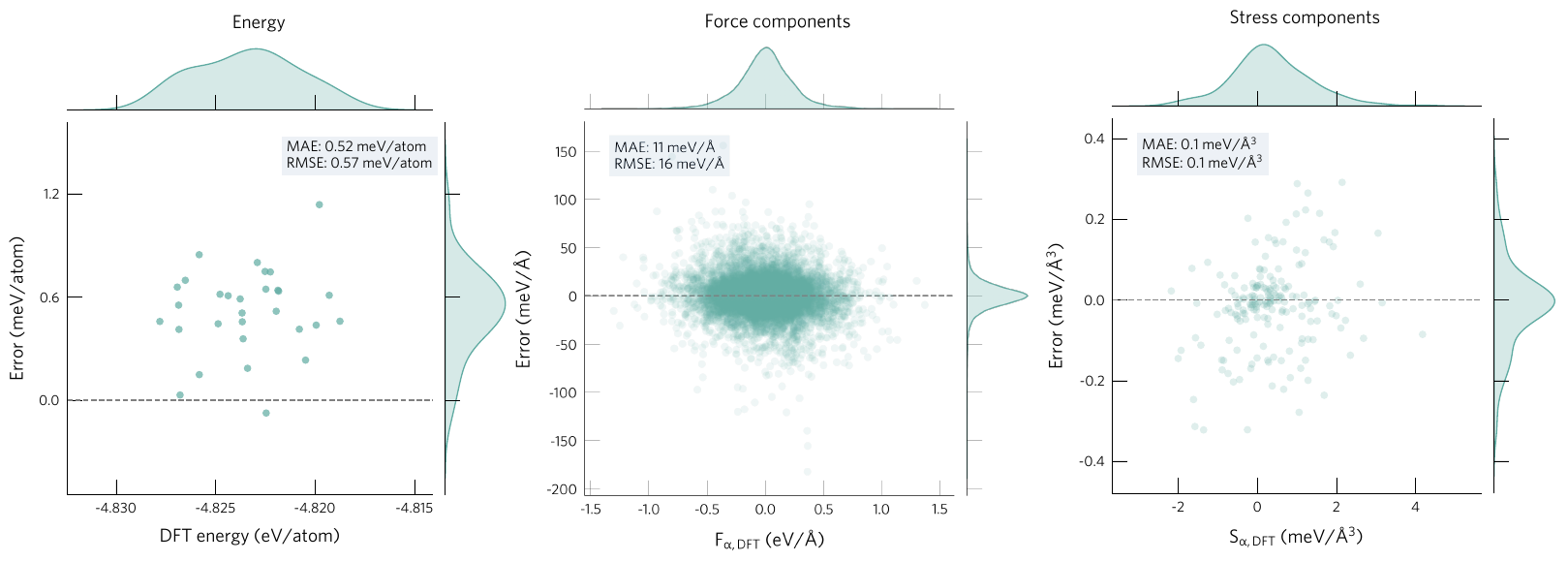}
    \caption{Distribution of absolute errors in the predicted energies, forces, and stresses for the test set of \kvc{V}{Cl}{0}. These defect structures were based on 159-atom supercells.}
    \label{sfig:errors_neutral}
\end{figure}

\begin{figure}[ht]
    \centering
    \includegraphics[width=0.9\linewidth]{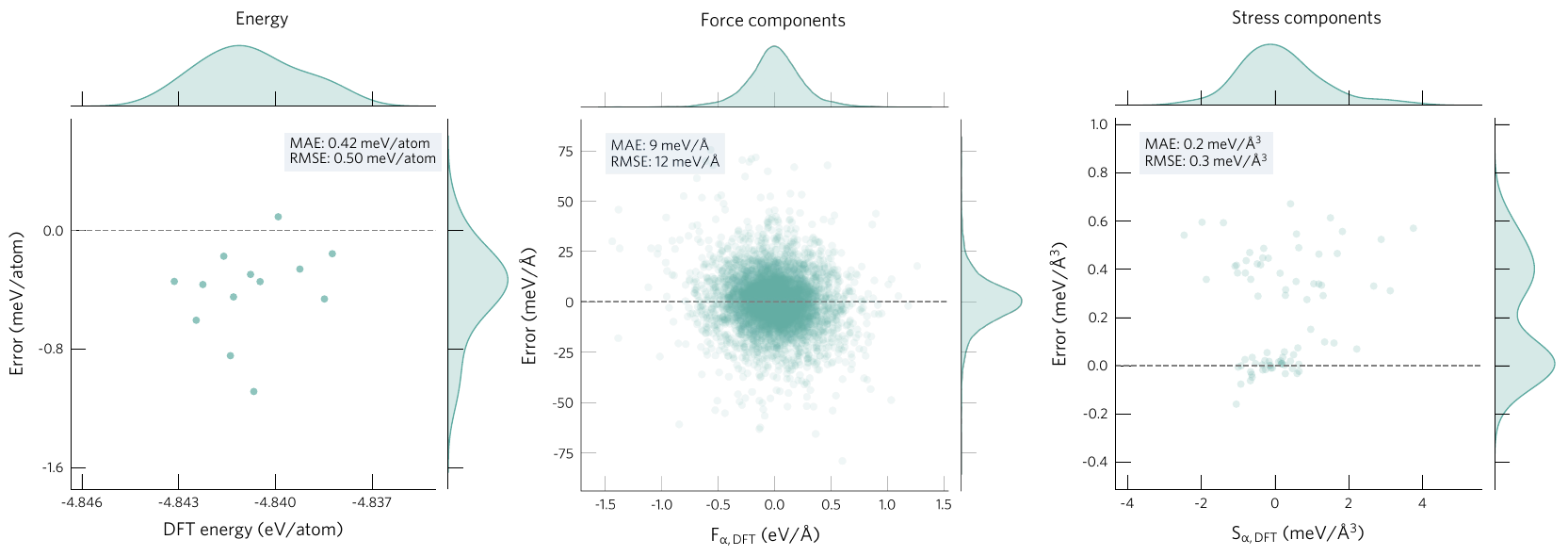}
    \caption{Distribution of absolute errors in the predicted energies, forces, and stresses for the test set of \kvc{V}{Cl}{+1}.}
    \label{sfig:errors_plus1}
\end{figure}

\begin{figure}[ht]
    \centering
    \includegraphics[width=0.4\linewidth]{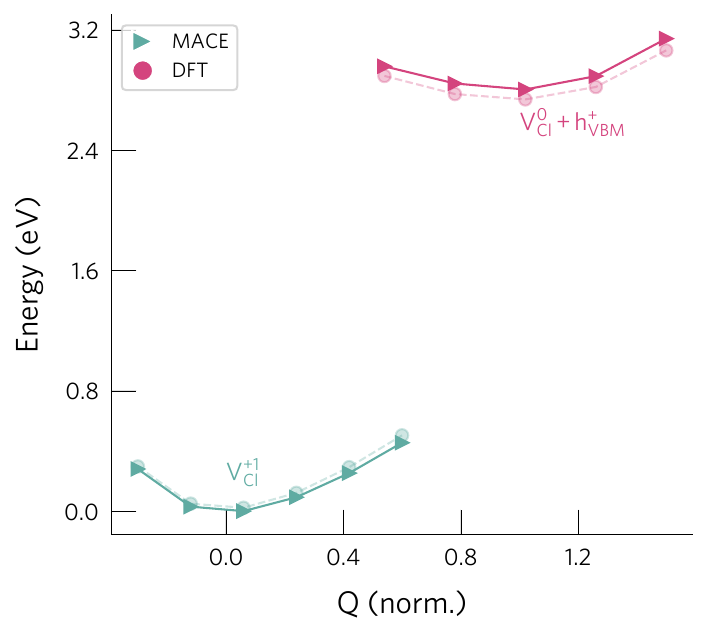}
    \caption{Comparison of predicted energies by DFT (HSE+SOC) and the MACE model for the \SI{0}{K} potential energy surfaces of \kvc{V}{Cl}{+1} and \kvc{V}{Cl}{0}. The structures were generated through a linear interpolation between the ground-state configurations of each charge state.}
    \label{sfig:errors_cc_diagram}
\end{figure}

We validated the performance of our machine learning force fields by comparing their predicted energies, forces and stresses on a test set of configurations with the ground-truth DFT values. As shown in Figs. \labelcref{sfig:errors_neutral,sfig:errors_plus1}, the predicted properties exhibit small absolute errors with no outliers, demonstrating that the models accurately describe the potential energy surfaces of each system at \SI{300}{K}. Carefully analysing the distribution of the errors is key to assessing the accuracy of the models for the defective systems, which have more complex energy landscapes than pristine crystals and, as a result, are more challenging to describe. Beyond validating the models on configurations sampled from MD, we also validated them on the \SI{0}{K} potential energy surfaces, which we generated by performing a linear interpolation between the ground-state structures of each charge state (Fig.~\ref{sfig:errors_cc_diagram}).

\begin{figure}[ht]
    \centering
    \includegraphics[width=0.55\linewidth]{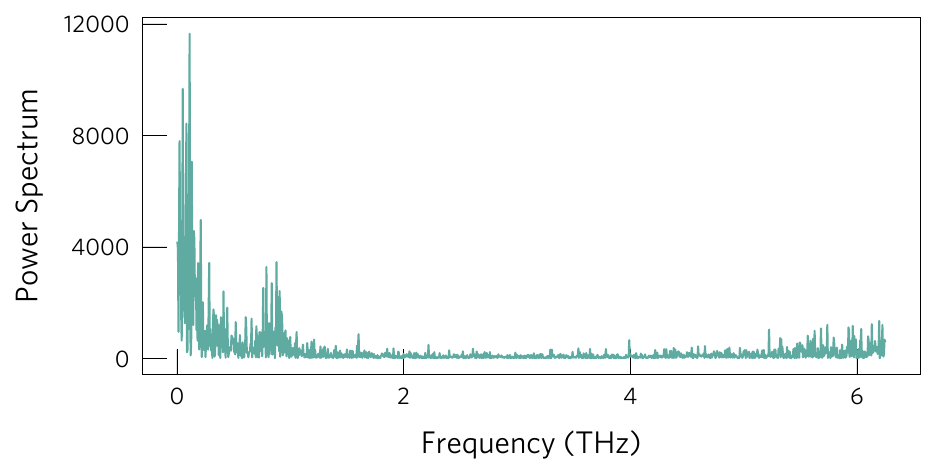}
    \caption{Fourier transform of the observed optical level during the molecular dynamics trajectory of \kvc{V}{Cl}{0}, illustrating that the strongest peaks occur for low frequencies ($\approx$\SI{0.12}{THz}).
    }
    \label{sfig:fourier_transform}
\end{figure}

\begin{figure}[ht]
    \centering
    \includegraphics[width=0.45\linewidth]{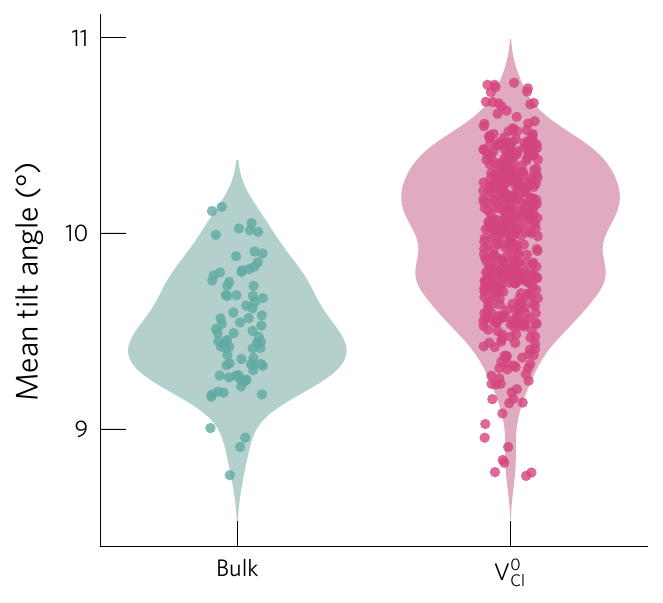}
    \caption{Mean tilt angles of bulk \ce{CsPbCl3} and \kvc{V}{Cl}{0}, in green and pink, respectively, during an NPT trajectory at \SI{300}{K}, illustrating the higher tilt angles of the supercell containing the Cl vacancy.
    }
    \label{sfig:tilt_angles}
\end{figure}

\clearpage

\bibliographystyle{rsc} 
\bibliography{biblio}
\end{document}